\documentclass[final,5p,times,twocolumn]{elsarticle}

\usepackage{graphicx}
\usepackage{dcolumn}
\usepackage{bm}
\usepackage{amssymb}

\journal{Journal of Physics and Chemistry of Solids}

\begin{document}

\begin{frontmatter}



\title{Multiple superconducting gap and anisotropic spin fluctuations in iron arsenides: Comparison with nickel analog}


\author[1]{Z. Li}
\author[2]{S. Kawasaki}
\author[2]{T. Oka}
\author[2]{T. Tabuchi}
\author[2]{Y. Ooe}
\author[2]{M. Ichioka}
\author[1]{Z. A. Ren}
\author[1]{Z. X. Zhao}
\author[1]{J. L. Luo}
\author[1]{N. L. Wang}
\author[1]{X. C. Wang}
\author[1]{Q. Q. Liu}
\author[1]{C. Q. Jin}
\author[3]{C.T. Lin}
\author[1,2]{Guo-qing Zheng} 

\address[1]{Institute of Physics and Beijing National Lab for Condensed Matter Physics, Chinese Academy of Science, Beijing, 100190, China }
\address[2]{Department of Physics, Okayama University, Okayama 700-8530, Japan}
\address[3]{Max Planck Institute, Heisenbergstrasse 1, D-70569 Stuttgart, Germany}

\begin{abstract}
We present extensive $^{75}$As NMR and NQR data on the
superconducting arsenides PrFeAs$_{0.89}$F$_{0.11}$ ($T_c$=45 K),
LaFeAsO$_{0.92}$F$_{0.08}$ ($T_c$=27 K), LiFeAs ($T_{\rm c}$ = 17
K) and Ba$_{0.72}$K$_{0.28}$Fe$_2$As$_{2}$ ($T_{\rm c}$ = 31.5 K)
single crystal, and compare with the nickel analog
LaNiAsO$_{0.9}$F$_{0.1}$ ($T_c$=4.0 K) . In contrast to
LaNiAsO$_{0.9}$F$_{0.1}$ where the superconducting gap is shown to
be isotropic,  the spin lattice relaxation rate $1/T_1$ in the
Fe-arsenides decreases below $T_{\rm c}$ with no coherence peak and
shows a step-wise variation at low temperatures. The Knight shift
decreases below $T_c$ and shows a step-wise $T$ variation as well.
These results indicate spin-singlet superconductivity with
multiple gaps in the Fe-arsenides. The Fe antiferromagnetic spin
fluctuations are anisotropic and weaker compared to underdoped
copper-oxides or cobalt-oxide superconductors, while there is no
significant electron correlations in LaNiAsO$_{0.9}$F$_{0.1}$. We
will discuss the implications of these results and highlight the
importance of the Fermi surface topology.
\end{abstract}


\end{frontmatter}



\section{Introduction}
\label{}

The recent discovery of superconductivity in LaFeAsO$_{1-x}$F$_x$
at the  transition temperature $T_{\rm c}$ = 26 K \cite{Kamihara}
has attracted great attention. Soon after the initial work,
$T_{\rm c}$ was raised to 55 K in SmFeAsO$_{1-x}$F$_x$
\cite{Ren3}, which is the highest among materials except cuprates.
These compounds have a ZrCuSiAs type structure (P4/nmm)  in which
FeAs forms a two-dimensional network similar to the CuO$_2$
plane in the cuprates case. By replacing O with F, {\it electrons}
are doped. After the discovery of ReFeAsO (Re: rare earth,
so-called 1111 compound), several other Fe-pnictides have been
found to superconduct. BaFe$_2$As$_2$ (so-called 122 compound) has
a ThCr$_2$Si$_2$-type structure. By replacing Ba with K, {\it
holes} are doped and $T_c$ can be as high as 38 K \cite{Rotter}.
Another arsenide, LiFeAs (so-called 111 compound) which has
Cu$_{2}$Sb type tetragonal structure, was discovered to show
superconductivity even in stoichiometric composition\cite{XCWang}.
The common feature of these three systems are the iron arsenide
plane which dominates the properties of these compounds and hosts the superconductivity.

We have used nuclear magnetic resonance (NMR) and nuclear
quadrupole resonance (NQR) techniques to study the pairing
symmetry and spin fluctuations in the normal state. We find that
these supercondutors are in the spin-singlet state with multiple
gaps, and the latter property is quite different from the cuprate
case. The antiferromagnetic spin fluctuation is weaker than the
cuprates and are anisotropic in the spin space.

\section{Experiments}

The preparation of the samples of PrFeAs$_{0.89}$F$_{0.11}$
($T_c$=45 K), LaFeAsO$_{0.92}$F$_{0.08}$ ($T_c$=23 K),
LiFeAs($T_{\rm c}$ = 17 K), LaNiAsO$_{0.9}$F$_{0.1}$ ($T_{\rm c}$
= 4.0 K), and single crystal Ba$_{0.72}$K$_{0.28}$Fe$_2$As$_{2}$
($T_{\rm c}$ = 31.5 K) are published elsewhere \cite{Ren, ChenPRL,
XCWang, Li, Lin}. NQR and NMR measurements were carried out by
using a phase coherent spectrometer. The NMR spectra were taken by
sweeping the magnetic field at a fixed frequency. The spin-lattice
relaxation rate $1/T_{1}$ was measured by using a single
saturation pulse.

\section{Results and Discussion}
\subsection{Knight shift}

\begin{figure}[h]
\includegraphics[width=6.5cm]{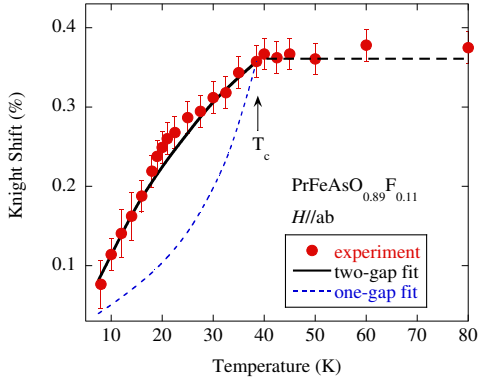}
\caption{\label{figPrK} (color online) The temperature variation
of $^{75}$As Knight shift of PrFeAs$_{0.89}$F$_{0.11}$ with
$H\parallel ab$. The solid curve is a fitting of two $d$-wave gaps
with  $\Delta_1(T=0) = 3.5k_BT_C$ and a relative weight of 0.4,
and $\Delta_2 (T=0) = 1.1k_BT_C$ with a relative weight of 0.6
(see text). The broken curve below $T_{c}$ is a simulation for the
larger gap alone. }
\end{figure}

\begin{figure}[h]
\includegraphics[width=6.5cm]{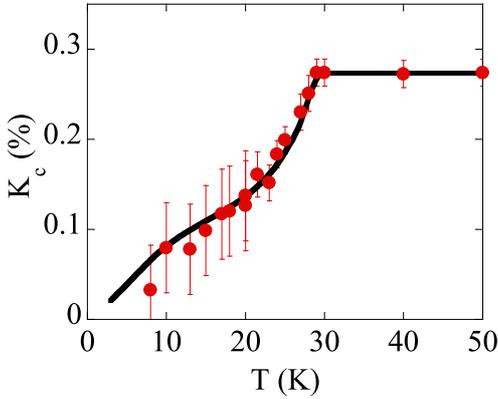}
\caption{\label{figBaK} (color online) The Knight shift data of
Ba$_{0.72}$K$_{0.28}$Fe$_2$As$_{2}$ with $H \parallel c$-axis. The
arrow indicates $T_{c}$. The curve below $T_{c}$ is a fit to a
two-gap model (see text). }
\end{figure}

Figure \ref{figPrK} shows the temperature dependence of the Knight
shift for PrFeAs$_{0.89}$F$_{0.11}$ ($T_c$=45 K) with the magnetic
field ($H$) applied along the $ab$-direction \cite{Matano}. The shift
decreases below $T_c$ and goes to zero at the $T$=0 limit.
The Knight
shift of Ba$_{0.72}$K$_{0.28}$Fe$_2$As$_{2}$ with $H$ parallel to the $c$-axis \cite{Matano_Ba} also
decreases below $T_c$, as seen in  Fig. \ref{figBaK}. The behavior is quite similar to the PrFeAs$_{0.89}$F$_{0.11}$ case.
These results indicate spin-singlet superconductivity.

However, the detailed $T$ variation of the Knight shift  is
different from that seen in usual spin-singlet superconductors
such as copper-oxides, where $K$ decreases rapidly below $T_{c}$
which is followed by a milder decrease at low temperatures, as
illustrated by the broken curve in Fig. \ref{figPrK}.  In
contrast, the decrease of the Knight shift shows a step-wise
behavior at a temperature about  half the $T_c$.

\subsection{$T_{1}$ in the superconducting state}

\begin{figure}[h]
\includegraphics[width=6.5cm]{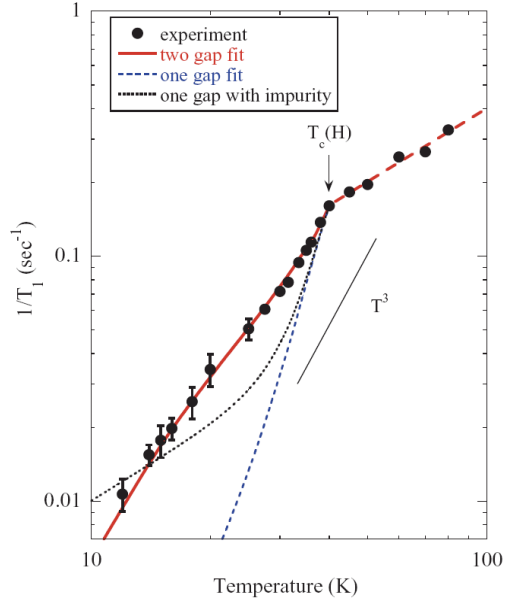}
\caption{\label{figPrT1} (color online) The temperature dependence of $^{19}$F spin-lattice relaxation rate
$1/T_{1}$ in PrFeAs$_{0.89}$F$_{0.11}$ measured at $\mu_{0} H =1.375$ T. The broken line indicates a relation
of $T_{1}T=$const which holds for a weakly correlated electron system. The thin straight line illustrates the
temperature dependence of $T^{3}$}
\end{figure}

\begin{figure}[h]
\includegraphics[width=6.5cm]{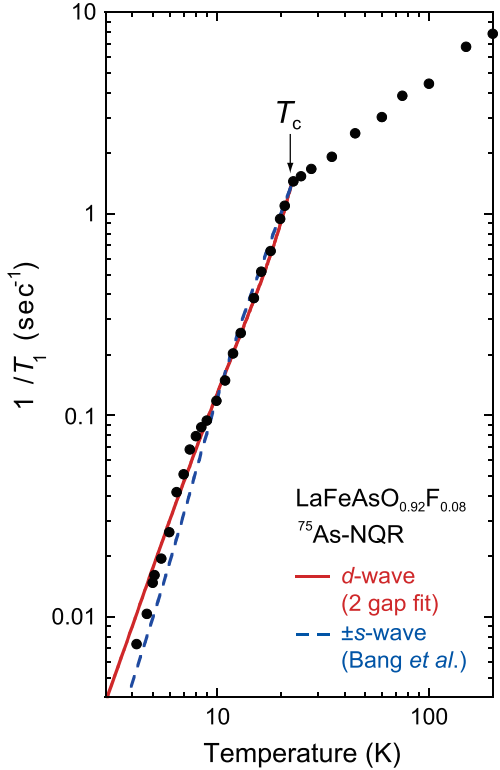}
\caption{\label{figLaT1} (color online)  $^{75}$As ($1/T_1$) in
LaFeAsO$_{0.92}$F$_{0.08}$. The solid curve is a two gap fit
assuming a $d$-wave symmetry with  parameters, $\Delta_1(0) = 4.2
k_{\rm B}T_{\rm c}$, $\Delta_2(0) = 1.6 k_{\rm B}T_{\rm c}$,  and
$\kappa = 0.6$ (see text). The dotted curve is a simulation
assuming two $s$-wave gaps that change signs with  parameters,
$\Delta_1(0) = 3.75 k_{\rm B}T_{\rm c}$, $\Delta_2(0) = 1.5 k_{\rm
B}T_{\rm c}$, and  $\kappa = 0.38$. }
\end{figure}

The step-wise decrease of the Knight shift is also
reflected in the temperature dependence of the $^{19}$F
spin-lattice relaxation rate $1/T_1$ \cite{Matano}, and is also
seen in LaFeAsO$_{0.92}$F$_{0.08}$ ($T_c$=23 K)\cite{Kawasaki} and
the hole-doped Ba$_{0.72}$K$_{0.28}$Fe$_2$As$_{2}$ ($T_c$=31.5
K)\cite{Matano_Ba}.
Figure \ref{figPrT1} shows the temperature dependence of $1/T_1$ measured by $^{19}$F NMR in
PrFeAs$_{0.89}$F$_{0.11}$ ($T_{c}=$ 45 K), and Fig. \ref{figLaT1} shows the temperature dependence of $1/T_1$
measured by $^{75}$As NQR in LaFeAsO$_{0.92}$F$_{0.08}$ ($T_c$=23 K). Below $T_c$, there is no coherence peak
for both samples. Moreover, a bending feature is seen around $T\sim T_c/2$ \cite{Matano, Kawasaki}. This
behavior is not expected in a single-gap superconductor.

\begin{figure}[h]
\includegraphics[width=6.5cm]{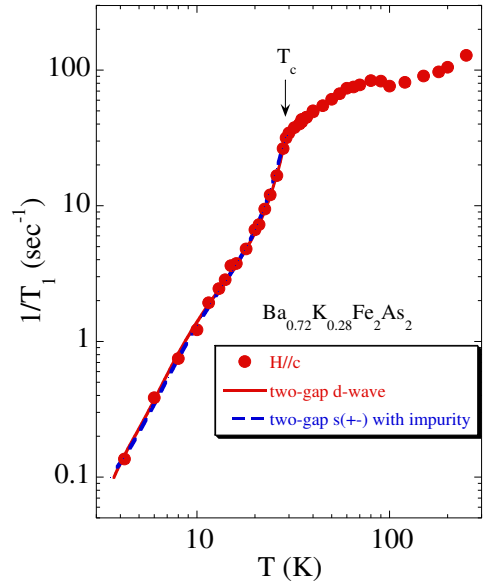}
\caption{\label{figBaT1} (color online) $T$-dependence of 1/$T_1$
in Ba$_{0.72}$K$_{0.28}$Fe$_2$As$_{2}$. The curves below $T_c$
(indicated by the arrow) are fits to  two-gap models  (see text).
}
\end{figure}

Figure \ref{figBaT1} shows the temperature dependence of $1/T_1$ measured by $^{75}$As NMR with the magnetic
field applied along the $c$-axis in Ba$_{0.72}$K$_{0.28}$Fe$_2$As$_{2}$ ($T_{\rm c}$ = 31.5 K) single crystal
\cite{Matano_Ba}. $1/T_1$ also shows a "knee"-shape around $T \sim T_{c}/2$. Namely, the sharp drop of $1/T_1$
just below $T_{\rm c}$ is gradually replaced by a slower change below $T\sim$ 15 K, then followed by another
steeper drop below. This "convex" shape is clearly different from the case of superconductors with a single gap
which shows a "concave" shape of $T$-variation.

We find that a two-gap model can reproduce the step-wise $T$
variation of $1/T_1$ and the Knight shift. The underlying physics
is that the system
is dominantly governed by a larger gap for $T$ near $T_c$ while at
sufficiently low $T$  it starts to "notice" the existence of a
smaller gap, resulting in another drop in $1/T_1$ below $T\sim$
$T_{c}/2$.
%
In the $d$-wave case with two gaps, where the density of states
(DOS) is $N_{s,i}(E)$ =
$N_{0,i}$$\frac{E}{\sqrt{E^2-\Delta_i^2}}$, the Knight shift and $1/T_{1s}$ in the
superconducting state are written as
\begin{eqnarray}
 \frac{K_s}{K_N}= \int N_s(E)\frac{\partial f (E)}{\partial E}dE
 \end{eqnarray}
\begin{eqnarray*}
\frac{T_{1N}}{T_{1s}}= \sum{ \frac{2}{k_BT} \int \int N_{s,i}(E)N_{s,i}(E')}
\end{eqnarray*}
\begin{eqnarray}
& & f(E)\left[ 1-f(E') \right] \delta(E-E')dEdE'
\end{eqnarray}
where $f(E)$ is the Fermi distribution function.
We find that the parameters 2$\Delta_1(0) = 7.0 k_{\rm B}T_{\rm
c}$, 2$\Delta_2(0) = 2.2 k_{\rm B}T_{\rm c}$  and  $\kappa =
0.4$ can fit the data of both the shift and $1/T_1$ very well, as shown by the solid curves in
Fig. 1 and Fig. 3, where
\begin{eqnarray}
\kappa = \frac{N_{0,1}}{N_{0,1}+N_{0,2}}
\end{eqnarray}
is the relative  DOS of the band(s) with  larger gap to the total
DOS.

Application of the same model to LaFeAsO$_{0.92}$F$_{0.08}$ gives
2$\Delta_1(0) = 8.4 k_{\rm B}T_{\rm c}$, 2$\Delta_2(0) = 3.2
k_{\rm B}T_{\rm c}$,  and  $\kappa = 0.6$ \cite{Kawasaki}. On the
other hand, for  Ba$_{0.72}$K$_{0.28}$Fe$_2$As$_{2}$,
$2\Delta_1(0)$ = 9.0 $k_{\rm B}T_{\rm c}$, $2\Delta_2(0)$ =
1.62$k_{\rm B}T_{\rm c}$, and $\kappa$ = 0.69  was obtained
\cite{Matano_Ba}. The same model can also fit the Knight shift
data, as seen in Fig. 2.


For the case of $s^{\pm}$-gap \cite{Mazin,Kuroki}, recent
calculations have shown that scattering between the  different
bands may reduce the coherence peak just below $T_{\rm c}$
\cite{Chubukov,Bang}. Following Ref.\cite{Bang}, we calculated
$1/T_1$ for the $s^{\pm}$-gap model, by introducing the impurity
scattering parameter $\eta$ in the energy spectrum,
$E=\omega+i\eta$. The parameters 2$\Delta_1(0) = 7.5 k_{\rm
B}T_{\rm c}$, 2$\Delta_2(0) = 3.0 k_{\rm B}T_{\rm c}$, $\kappa =
0.38$ and $\eta$=0.15$k_{\rm B}$$T_{\rm c}$, can well fit the data,
as shown in  Fig. 4. where
\begin{eqnarray}
\eta=\frac{\pi n_{\rm imp} (N_{1}+N_{2}) V^{2}}{1+\pi ^{2}
(N_{1}+N_{2})^{2} V^{2}}
\end{eqnarray}
In the equation, $n_{\rm imp}$ is the impurity concentration and
$V$ is the scattering potential at the impurity.
A similar set of parameters (2$\Delta_1(0) = 7.2 k_{\rm
B}T_{\rm c}$, 2$\Delta_2(0) = 1.68 k_{\rm B}T_{\rm c}$, $\kappa =
0.6$ and $\eta$=0.22$k_{\rm B}$$T_{\rm c}$) can fit the data of Ba$_{0.72}$K$_{0.28}$Fe$_2$As$_{2}$, see Fig.
\ref{figBaT1}.

\begin{figure}[h]
\includegraphics[width=6.5cm]{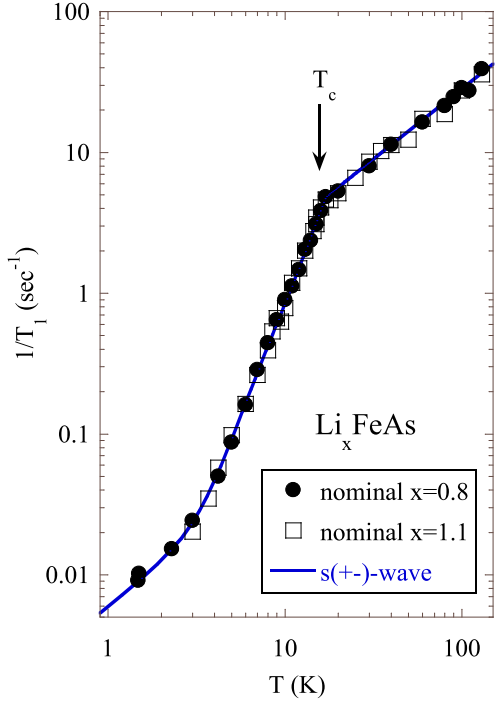}
\caption{\label{figLiT1} (color online) The $T$-dependence of
$1/T_{1}$ measured by NQR for Li$_{0.8}$FeAs and Li$_{1.1}$FeAs.
The curves below $T_{\rm{c}}$  are fits to the $s^{\rm \pm}$-wave
model with $\Delta_{1}^{+}=3.0$ $k_{\rm{B}} T_{\rm{c}}$,
$\Delta_{2}^{-}=1.3$ $k_{\rm{B}} T_{\rm{c}}$, and the impurity
scattering rate  $\eta = 0.26$ $k_{\rm B}T_{\rm c}$ (see text).}
\end{figure}

The results for LiFeAs\cite{ZLi} (Fig. \ref{figLiT1}) are shown in
Fig. 6.  We measured two Li$_{x}$FeAs samples with nominal $x=0.8$
and $1.1$. The physical properties including the NMR results are
the same. This supports that only stoichiometric compound can be
formed.\cite{JHTapp} $1/T_1$ below $T_c$ shows a qualitatively
similar behavior as the previous three samples, but the behavior
at low temperatures is a little different. Namely, $1/T_1$ becomes
to be proportional to $T$  below $T\leq T_c$/4, which indicates
that a finite DOS is induced by the impurity scattering.
Obviously, this would occur in a d-wave case. On the other hand,
it is also possible
 in the $s^{\pm}$ case provided that the scattering between the electron- and hole-pocket is strong. Calculation by the  $s^{\pm}$-wave model shows that the gap value of LiFeAs is smaller than other compounds, but
the impurity scattering is much larger ($\eta=0.26 k_BT_c$) \cite{ZLi}.
The gap parameters for all Fe-arsenide samples are summarized
in Tab. \ref{tab:table1}.
To summarize,  the multiple gap feature is universal to all Fe-arsenids, which
probably associated with the multiple electronic bands \cite{Singh}.

\begin{figure}[h]
\includegraphics[width=7.5cm]{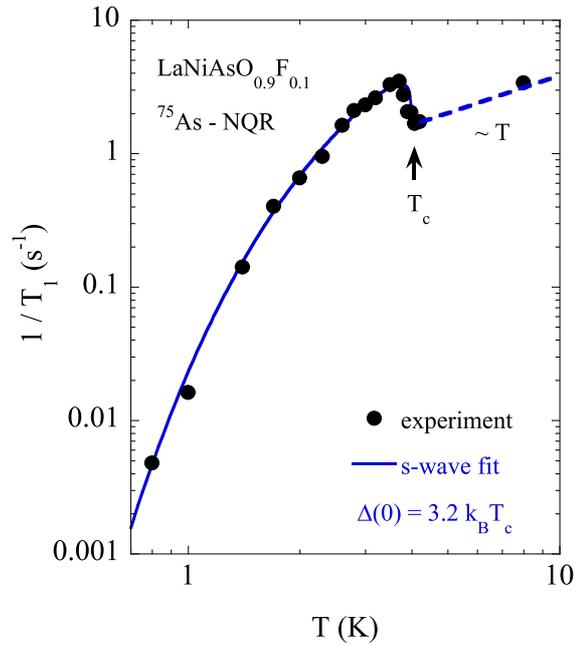}
\caption{\label{figNiT1} (color online) The $T$ dependence of the
spin lattice relaxation rate, $1/T_1$, for
LaNiAsO$_{0.9}$F$_{0.1}$. The arrows indicate $T_c$. The broken
straight lines show the relation $1/T_1 \propto T$, and the curves
below $T_c$ are fits to the BCS model with the gap size indicated
in the figure.}
\end{figure}

By strong contrast, the nickel analog of LaFeAsO$_{1-x}$F$_{x}$,
namely, LaNiAsO$_{0.9}$F$_{0.1}$ has different
behavior.\cite{Tabuchi} As seen in Fig. \ref{figNiT1}, $1/T_{1}$
shows a well-defined coherence peak just below $T_{c}$, which is a
finger print of superconductors with an isotropic gap. This is in
sharp contrast to various Fe-arsenides reported so
far\cite{Matano,Kawasaki,Matano_Ba,Ishida,Grafe,Fukazawa}. At low
temperatures, $1/T_1$ decreases as an exponential function of $T$.
The solid curves in Fig. \ref{figNiT1} are calculations using the BCS model.
Following Hebel \cite{Hebel}, we convolute  $N_{s}(E)$ with a
broadening function $B(E)$ which is approximated with a
rectangular function centered at $E$ with a height of $1/2\delta$.
The solid curves below $T_c$ shown in Fig. \ref{figNiT1} is
calculation with 2$\Delta(0)=3.2 k_BT_c$ and
$r\equiv\Delta(0)/\delta$=5. Such $T$-dependence of $1/T_1$ in the
superconducting state is in striking contrast to that for
Fe-arsenide where no coherence peak was observed and the
$T$-dependence at low-$T$ does not show an  exponential behavior.
The striking difference may be ascribed to the different topology
of the Fermi surfaces. For Fe-arsenides, it has been proposed that
$d$-wave \cite{Graser,Kuroki2} or sign reversal  $s$-wave gap
\cite{Mazin,Kuroki} can be stabilized  due to nesting by  the
connecting wave vector $Q=(\pi, 0)$. In LaNiAsO$_{0.9}$F$_{0.1}$,
however, there is no such Fermi surface nesting \cite{Xu}, and
thus the mechanism for the proposed gap symmetry does not exist.
Given that the $T_c$ is much lower in LaNiAsO$_{1-x}$F$_{x}$, our
result suggests the important role of the Fermi-surface topology
in the superconductivity of Fe-arsenides.

\begin{figure}[h]
\includegraphics[width=7.5cm]{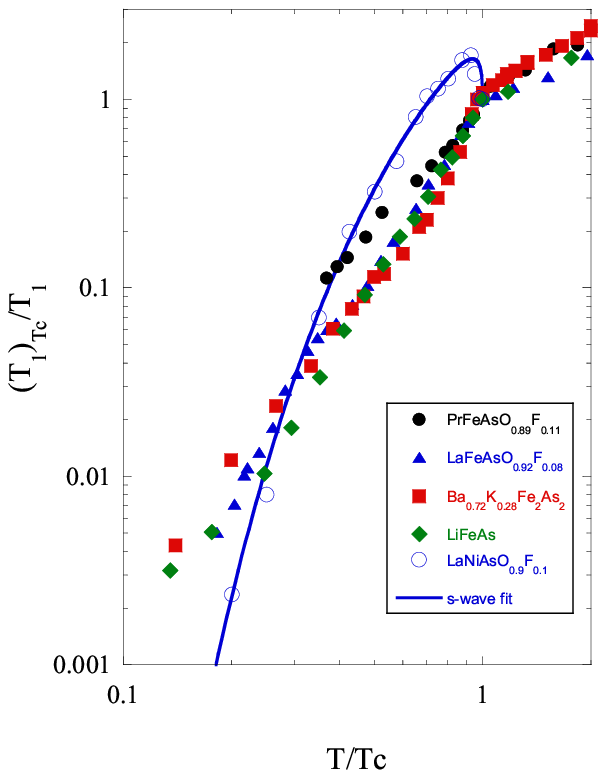}
\caption{\label{figT1} (color online) The normalized T-dependence
of $1/T_{1}$ for PrFeAs$_{0.89}$F$_{0.11}$ ($T_c$=45 K),
LaFeAsO$_{0.92}$F$_{0.08}$ ($T_c$=23
K),Ba$_{0.72}$K$_{0.28}$Fe$_2$As$_{2}$ ($T_{\rm c}$ = 31.5 K).
LiFeAs($T_{\rm c}$ = 17 K), LaNiAsO$_{0.9}$F$_{0.1}$ ($T_{\rm c}$
= 4.0 K).}
\end{figure}

Finally, for comparison, $1/T_{1}$ normalized by the value at $T_{c}$ are
shown in Fig. \ref{figT1} as a function of reduced temperature for all samples.

\begin{table*}
\caption{\label{tab:table1}The gap parameters. $\kappa =
N_{1}/(N_{1}+N_{2})$, $\eta=\frac{\pi n_{\rm imp} (N_{1}+N_{2})
V^{2}}{1+\pi ^{2} (N_{1}+N_{2})^{2} V^{2}}$, where $n_{\rm imp}$
is the impurity concentration and $V$ is the scattering potential
at the impurity.}
\begin{tabular}{cccccccc}
  \hline\hline
    & \multicolumn{3}{c}{$d$-wave} &
    \multicolumn{4}{c}{$s^{\pm}$-wave}\\
    &$\Delta_{1}$/$k_{\rm B}T_{c}$ &$\Delta_{2}$/$k_{\rm B}T_{c}$ &$\kappa$
    & $\Delta_{1}$/$k_{\rm B}T_{c}$ &$\Delta_{2}$/$k_{\rm B}T_{c}$ & $\kappa$ &$\eta$/$k_{\rm B}T_{c}$\\
  \hline
   PrFeAsO$_{0.89}$F$_{0.11}$ & 3.5 & 1.1 & 0.4 & & & & \\
   LaFeAsO$_{0.92}$F$_{0.08}$ & 4.2 & 1.6 & 0.6 & 3.75 & 1.5 & 0.38 & 0.15 \\
   Ba$_{0.72}$K$_{0.28}$Fe$_{2}$As$_{2}$ & 4.5 & 0.81 & 0.69 & 3.6 & 0.84 & 0.6 & 0.22 \\
   LiFeAs & & & & 3.0 & 1.3 & 0.5 & 0.26 \\
  \hline\hline
\end{tabular}
\end{table*}

\subsection{Normal state properties}

\begin{figure}[h]
\includegraphics[width=7.5cm]{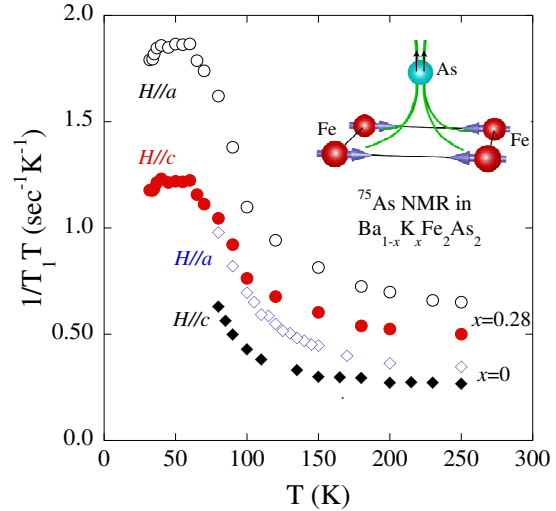}
\caption{\label{figBaT1T} (color online) The quantity
$^{75}$($1/T_1T$) in the normal state of
Ba$_{0.72}$K$_{0.28}$Fe$_2$As$_{2}$ (circles) and in the
paramagnetic state of BaFe$_2$As$_{2}$ (diamonds). The arrows in
the inset  illustrate  the larger component of the  fluctuating
field of Fe  and that seen by the As site.}
\end{figure}

Next, we discuss on the character of spin fluctuations. Figure \ref{figBaT1T} shows the temperature variation
of $1/T_1T$ in Ba$_{0.72}$K$_{0.28}$Fe$_2$As$_2$. One notices that, in the normal state above $T_c$, $1/T_1T$
increases with decreasing $T$, which is an indication of antiferromagnetic electron correlation, since both
$K^a$ and $K^c$ slightly decrease with decreasing $T$, but becomes a constant below $T\sim$ 70 K
\cite{Matano_Ba}, which resembles closely the cuprate \cite{Zheng-LaPr} or cobaltate cases \cite{Zheng-Co}.

\begin{figure}[h]
\includegraphics[width=8.5cm]{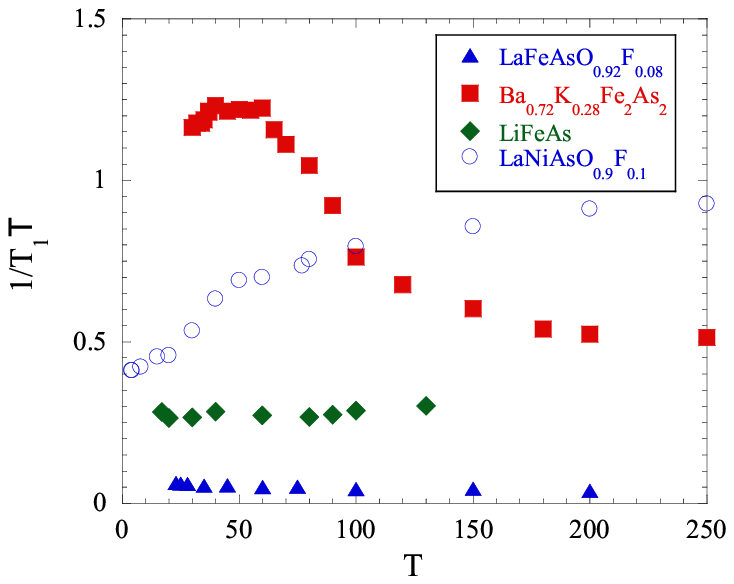}
\caption{\label{figT1T} (color online) The  T-dependence of
$1/T_{1}T$ for LaFeAsO$_{0.92}$F$_{0.08}$ ($T_c$=23
K),Ba$_{0.72}$K$_{0.28}$Fe$_2$As$_{2}$ ($T_{\rm c}$ = 31.5 K),
LiFeAs($T_{\rm c}$ = 17 K), LaNiAsO$_{0.9}$F$_{0.1}$ ($T_{\rm c}$
= 4.0 K).}
\end{figure}

Figure \ref{figT1T} compares $1/T_{1}T$ for four samples. The data for
LaFeAsO$_{0.92}$F$_{0.08}$, LiFeAs, LaNiAsO$_{0.9}$F$_{0.1}$ are
measured by NQR, which correspond to $H \parallel c$-axis, since
the principle axis of the NQR tensors is along the $c$-axis.
The data for Ba$_{0.72}$K$_{0.28}$Fe$_2$As$_{2}$ is measured by NMR with
$H \parallel c$-axis. The $1/T_{1}T$ of hole-doped
Ba$_{0.72}$K$_{0.28}$Fe$_2$As$_{2}$ increase with decreasing
temperature as discussed above. The electron-doped
LaFeAsO$_{0.92}$F$_{0.08}$ also show similar behavior, although
the increasing is very small. The $1/T_{1}T$ of  stoichiometric
LiFeAs is almost constant. While $1/T_{1}T$ of
LaNiAsO$_{0.9}$F$_{0.1}$ decrease with decreasing temperature.
These results suggest that the antiferromagnetic spin fluctuations are
stronger in Ba$_{0.72}$K$_{0.28}$Fe$_2$As$_{2}$ and
LaFeAsO$_{0.92}$F$_{0.08}$, but quite weak in   LiFeAs and
LaNiAsO$_{0.9}$F$_{0.1}$. This difference may be  understood by the difference of the Fermi
surface topology. There are not only hole-pockets and
electron-pockets but also nesting between them in
Ba$_{0.72}$K$_{0.28}$Fe$_2$As$_{2}$ and LaFeAsO$_{0.92}$F$_{0.08}$
,which can promote spin fluctuations. While in LiFeAs there is no
such nesting, although there are still hole-pockets and
electron-pockets. Lacking of such nesting make the spin
fluctuation become weaker than Ba$_{0.72}$K$_{0.28}$Fe$_2$As$_{2}$
and LaFeAsO$_{0.92}$F$_{0.08}$. In LaNiAsO$_{0.9}$F$_{0.1}$ there
is no hole-pockets, then nesting can not happen, therefore the
spin fluctuations are not expected.

\begin{figure}[h]
\includegraphics[width=7.5cm]{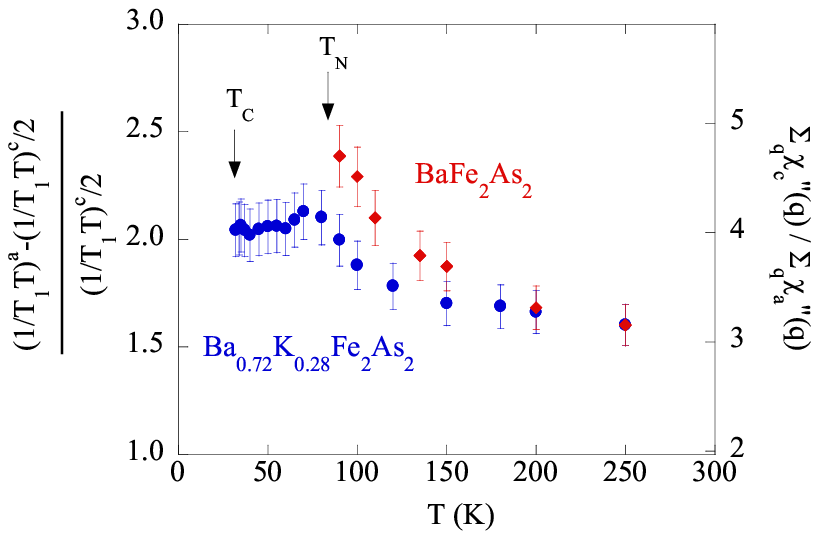}
\caption{\label{figR} (color online) The T dependence of the anisotropy of the spin fluctuations seen at the As
site in terms of $\frac{\sum_{\mathbf{q}}A_{hf}^{c}(\mathbf{q})^{2}\chiup_{c}^{\prime\prime}(\mathbf{q})}
{\sum_{\mathbf{q}}A_{hf}^{a}(\mathbf{q})^{2}\chiup_{a}^{\prime\prime}(\mathbf{q})}$ for the left axis and
$\frac{\sum_{\mathbf{q}}\chiup_{c}^{\prime\prime}(\mathbf{q})}{\sum_{\mathbf{q}}\chiup_{a}^{\prime\prime}(\mathbf{q})}$
for the right axis.}
\end{figure}

Finally, we discuss the anisotropy of the spin fluctuations. In a general form, $1/T_1T$  is written as
\begin{eqnarray} 
\frac{1}{T_1T}= \frac{\pi k_B \gamma^2_n }{(\gamma_e \hbar )^2}
\sum_q A_{hf}^2 \frac{\chi ''_{\perp}(q,\omega)}{\omega},
\end{eqnarray}
 where
$\chi ''_{\perp}(q,\omega)$ is the imaginary part of the dynamical
susceptibility  perpendicular to the applied field. The
anisotropic ratio in the form of
$\frac{\sum_{\mathbf{q}}A_{hf}^{c}(\mathbf{q})^{2}\chiup_{c}^{\prime\prime}(\mathbf{q})}
{\sum_{\mathbf{q}}A_{hf}^{a}(\mathbf{q})^{2}\chiup_{a}^{\prime\prime}(\mathbf{q})}$
for Ba$_{0.72}$K$_{0.28}$Fe$_2$As$_{2}$ and parent compound are
shown in Fig.\ref{figR}. The larger magnitude of $1/T_1T$ along
the $a$-axis direction than that along the $c$-axis direction
indicates that there are stronger fluctuations along the $c$-axis
direction seen by  the As-site. Neutron experiment found that, in
the undoped BaFe$_2$As$_2$ compound, the ordered Fe magnetic
moment is along the $a$-direction and forms a stripe
\cite{neutron}. Since the As atom sits in the position above
(below) the middle of four Fe-atoms, our result implies that, in
the Fe site, a stronger fluctuating fields exist along the
$a$-axis direction, as illustrated in the inset of Fig.
\ref{figBaT1T}. It is remarkable that the antiferromagnetic
fluctuations of Fe are anisotropic in spin space. Namely, $\chi
''_{\pm}(Q)$ is  much larger than $\chi ''_{zz}(Q)$, where $z$ is
along the $c$-axis direction and $Q$ is the spin fluctuation wave
vector. This is in contrast to the high-$T_c$ cuprates where the
spin fluctuations are believed to be isotropic, but similar
situation was encountered in cobaltate superconductor
\cite{Matano-2}. The relationship between the energy- and
$q$-dependence of the spin fluctuations (SF) and possible
SF-induced superconductivity has been studied both theoretically
\cite{Moriya} and experimentally \cite{ZhengTl}.  To our knowledge,
however, the relationship between the anisotropy of SF and
superconductivity  has  been less explored so far. We hope that
our results  will stimulate  more theoretical work in this regard.

\section{Conclusion}

In summary, we have presented the  NMR and NQR results on  the
electron-doped Fe-arsenides  PrFeAs$_{0.89}$F$_{0.11}$ ($T_c$ = 45 K),
LaFeAsO$_{0.92}$F$_{0.08}$ ($T_c$ = 23 K), stoichiometric LiFeAs($T_{\rm c}$ = 17 K), and the hole-doped Fe-arsenide
Ba$_{0.72}$K$_{0.28}$Fe$_{2}$As$_{2}$ ($T_{\rm c}$ = 31.5 K).
 We find there are
multiple gaps in iron arsenides where Knight shift and
$1/T_{1}$ does not follow a simple power-law nor exponential
function.
However,
the $1/T_{1}$ of the Nickel analog LaNiAsO$_{0.9}$F$_{0.1}$ shows a well-defined  coherence
peak just below $T_c$ and an exponential behavior at lower temperatures. These properties indicate that it is a conventional BCS
superconductor.
The difference between the Fe-arsenides and the Ni-analog may be understood by the differene of the Fermi surface topology, and therefore highlights the important role of
the Fermi-surface topology in pairing symmetry of the iron arsenides superconductors.

In the normal state, all iron arsenides show weak antiferromagnetic spin correlations. Our data also show that
the sample with weaker correlations has a lower $T_{c}$, and this may imply the $T_{c}$ has a relationship with the
structure of Fermi surface. Moreover, the spin fluctuations are anisotropic in spin space, which is different
from cuprates.

We gratefully acknowledge the support from CAS, National Science
Foundation of China,  JSPS and MEXT of Japan .




\begin{thebibliography}{00}

\bibitem{Kamihara}
Y. Kamihara {\it et al},
J. Am. Chem. Soc. {\bf130}, 3296 (2008).

\bibitem{Ren3}
Z. -A. Ren {\it et al},
Chin. Phys. Lett. {\bf25}, 2215 (2008).

\bibitem{Rotter}
M. Rotter {\it et al}, Phys. Rev. Lett. {\bf 101}, 107006 (2008).

\bibitem{XCWang}
X. C. Wang {\it et al},
Solid State Commun. {\bf 148}, 538
(2008).

\bibitem{Ren}
Z.-A. Ren  {\it et al},
Materials Research Innovations {\bf 12}, 105 (2008).

\bibitem{ChenPRL}
G. F. Chen {\it et al},
Phys. Rev. Lett. {\bf 101}, 057007 (2008).

\bibitem{Li}
Z. Li {\it et al},
Phys. Rev. B {\bf 78}, 060504(R) (2008).

\bibitem{Lin}
G.L. Sun {\it et al},
arXiv:0901.2728 (2009).

\bibitem{Matano}
K. Matano  {\it et al},
Europhys. Lett. {\bf 83}, 57001 (2008).

\bibitem{Matano_Ba}
K. Matano {\it et al},
Europhys. Lett. {\bf 87}, 27012 (2009).

\bibitem{Kawasaki}
S. Kawasaki {\it et al},
Phys. Rev. B {\bf 78}, 220506 (R) (2008).

\bibitem{Mazin}
I. I. Mazin {\it et al},  Phys. Rev. Lett. {\bf 101}, 057003
(2008).

\bibitem{Kuroki}
K. Kuroki {\it et al},
Phys. Rev. Lett. {\bf 101}, 087004 (2008).

\bibitem{Chubukov}
A.V. Chubukov {\it et al}, Phys. Rev. B {\bf 78}, 134512 (2008).
D. Parker {\it et al}, {\it ibid}, 134524 (2008). M. M. Parish
{\it et al},  {\it ibid}, 144514 (2008).

\bibitem{Bang}
Y. Bang and H. -Y. Choi, Phys. Rev. B {\bf 78}, 134523 (2008); Y.
Nagai {\it et al}, New J. Phys. {\bf 10}, 103026 (2008).

\bibitem{ZLi}
Z. Li {\it et al}
J. Phys. Soc. Jpn. {\bf 79}, 083702 (2010).

\bibitem{JHTapp}
X. C. Wang {\it et al},
Solid State Commun. 148, 538 (2008)
J. H. Tapp {\it et al},
Phys. Rev.\ B {\bf 78}, 060505 (2008).

\bibitem{Singh}
D.J. Singh and M. H. Du,
Phys. Rev. Lett. {\bf 100}, 237003
(2008).

\bibitem{Tabuchi}
T. Tabuchi {\it et al}, Phys. Rev. B {\bf 81}, 140509 (2010).

\bibitem{Ishida}
Y. Nakai {\it et al},
J. Phys. Soc. Jpn. {\bf 77}, 073701 (2008).

\bibitem{Grafe}
H.-J. Grafe {\it et al},
Phys. Rev. Lett. {\bf 101}, 047003 (2008).

\bibitem{Fukazawa}
H. Fukazawa {\it et al},
J. Phys. Soc. Jpn. {\bf 78}, 033704 (2009).

\bibitem{Hebel}
L. C. Hebel, Phys. Rev. {\bf 116}, 79 (1959).

\bibitem{Graser}
S. Graser {\it et al},
 New J. Phys. {\bf 11}, 025016  (2009).

\bibitem{Kuroki2}
K. Kuroki {\it et al},
Phys. Rev. B {\bf 79}, 224511 (2009).

\bibitem{Xu}
G. Xu {\it et al},
Europhys. Lett, {\bf 82}, 67002 (2008) .

\bibitem{Zheng-LaPr}
G.-q. Zheng {\it et al},
Phys. Rev. Lett. {\bf 90}, 197005 (2003).

\bibitem{Zheng-Co}
G. - q. Zheng {\it et al},
Phys. Rev. {\bf B 73}, 180503 (R) (2006).








\bibitem{neutron}
Q. Huang  {\it et al},
, Phys. Rev. Lett. {\bf 101}, 257003 (2008)

\bibitem{Matano-2}
K. Matano {\it et al},
Europhys. Lett. {\bf 84}, 57010 (2008).


\bibitem{Moriya}
T. Moriya and K. Ueda,  J. Phys. Soc. Jpn. {\bf 63}, 1871 (1994);
D. Monthoux and D. Pines, Phys. Rev. B {\bf 49}, 4261 (1994).

\bibitem{ZhengTl}
G.-q. Zheng {\it et al}, J. Phys. Soc. Jpn. {\bf 64}, 3184 (1995).



\end{thebibliography}
\end{document}